\documentclass[11pt]{article}
\usepackage[utf8]{inputenc}
\usepackage{amsmath}
\usepackage{amssymb}
\usepackage{graphicx}
\usepackage{hyperref}
\usepackage{cite}
\usepackage{booktabs}
\usepackage{geometry}
\title{Quantifying Energy and Cost Benefits of Hybrid Edge Cloud: Analysis of Traditional and Agentic Workloads
}
\author{Siavash Alamouti}

\date{}

\begin{document}

\maketitle

\begin{abstract}
\noindent This paper examines the workload distribution challenges in centralized cloud systems and demonstrates how Hybrid Edge Cloud (HEC) \cite{alamouti2022hec} mitigates these inefficiencies. Workloads in cloud environments often follow a Pareto distribution, where a small percentage of tasks consume most resources, leading to bottlenecks and energy inefficiencies. By analyzing both traditional workloads reflective of typical IoT and smart device usage and agentic workloads, such as those generated by AI agents, robotics, and autonomous systems, this study quantifies the energy and cost savings enabled by HEC. Our findings reveal that HEC achieves energy savings of up to \textbf{75\%} and cost reductions exceeding \textbf{80\%}, even in resource-intensive agentic scenarios. These results highlight the critical role of HEC in enabling scalable, cost-effective, and sustainable computing for the next generation of intelligent systems.
\end{abstract}

\section{INTRODUCTION}
The proliferation of IoT devices, AI agents, and robotics has redefined the nature of workloads in modern computing systems. With the emergence of optimized AI models and ongoing hardware advancements, most smart devices including smartphones, PCs, and IoT devices are already capable of running narrow AI models efficiently. While upcoming device upgrades will further enhance AI capabilities, current devices are sufficient for handling most inference workloads, making a device-first approach not only feasible but highly relevant for agentic workflows \cite{tensorflowlite}, \cite{jackson2019aws}.\\\\
These workloads are often \textbf{Pareto-distributed}  \cite{barroso2013datacenter}, \cite{xu2017cloudworkload}, \cite{reiss2011googlecluster},\cite{dean2008mapreduce}, \cite{calheiros2011cloudsim}, \cite{baliga2011energy}, \cite{tucker2018energy} where a small percentage of high-resource tasks dominate computational resources, while most tasks are lightweight. Centralized cloud systems, originally designed for web browsing and app-based transactions, struggle to meet the demands of dynamic, context-aware applications.\\\\
This paper explores the implications of HEC, which can process tasks locally on end devices when possible and offloads only high-resource tasks to the cloud or dedicated cloud gateways. To provide a comprehensive view, we analyze both \textbf{traditional workloads} which reflect typical smart devices with less intelligence and \textbf{agentic workloads} emerging in AI-driven systems like autonomous vehicles and robotics.

\section{ASSUMPTIONS AND PARAMETERS}
To quantify the benefits of  HEC compared to current centralized cloud approaches,  we have built mathematical models and simulations. To make sure we can quantify these benefits in a clear fashion, we make assumptions based on published analysis on the nature of workloads, energy consumption, bandwidth and cloud hosting costs. Below are the assumptions and justifications for them in our model.

\subsection{Data Generation}
\begin{itemize}
    \item \textbf{Traditional workloads:} 2.4 GB/day, translating to 876 GB/year per device.
    \item \textbf{Agentic workloads:} 20 GB/day, translating to 7,300 GB/year per device.
\end{itemize}

\subsection{Energy Consumption Assumptions}
\begin{itemize}
    \item Transmission of data to cloud nodes: 0.7 kWh/GB.
    \item Processing of data workloads on the cloud: 1.5 kWh/GB.
    \item Processing of data workloads on end devices: 0.5 kWh/GB.
\end{itemize}

\noindent The number used for energy consumption for the transmission to cloud are based on studies published in  \cite{masanet2020recalibrating}, \cite{venkatesh2021efficiency}. The numbers for cloud and device processing are based on results from \cite{akamai2020report}, \cite{cisco2021cloudindex}, and \cite{aws2023pricing}. As per these studies, the \textbf{3x difference in energy consumption between cloud and edge processing} is due to multiple factors beyond transmission energy. In cloud-based processing, devices must activate their communication interfaces (e.g., ethernet, Wi-Fi, LTE, or 5G) to send data to the cloud, which consumes additional energy at the device level. In device processing, most of this data remains local, eliminating the need for constant use of these communication interfaces and resulting in significant energy savings. Furthermore, cloud data centers require resource-intensive activities such as virtualization, workload orchestration, and redundancy mechanisms, which add computational and operational overheads. These are further compounded by the energy needed for cooling, infrastructure management, and large-scale storage systems. In contrast, edge processing directly utilizes the local context to minimize computational complexity and avoids these cloud-scale energy demands. Also, for many use cases, the context already available locally on these devices needs to be properly packaged and transmitted to the cloud. By reducing reliance on communication interfaces and leveraging lightweight, efficient device operations, processing workloads on devices, achieves considerable energy savings compared to centralized cloud architectures.
\subsection{Bandwidth and Hosting Cost assumptions:}
Bandwidth costs represent the price of transferring data to the cloud or across content delivery networks (CDNs), with reports indicating a range of $\$$0.01–$\$$0.12 per GB, depending on infrastructure and usage scale. Hosting costs, which include compute and storage expenses, are derived from cloud providers such as AWS, Google Cloud Platform, and Microsoft Azure. These costs average around $\$$0.20 per GB for typical dynamic workloads, particularly those requiring AI inference or high-throughput data processing. Together, these assumptions provide a realistic and balanced basis for estimating cloud costs in comparative analyses. In our model, we assume.
\begin{itemize}
    \item Bandwidth: \$0.10/GB.
    \item Hosting: \$0.20/GB.
\end{itemize}

\noindent These cost assumptions for bandwidth and hosting are reasonable and supported by industry data published in \cite{google2023pricing}, \cite{azure2023pricing}, \cite{hinton2015distilling}, \cite{mckinsey2022edgeai}, \cite{arm2023cortex}.

\subsection{Pareto Allocation assumption:}
The task allocation between end devices and the centralized cloud in this analysis is based on the Pareto principle, which suggests that a small fraction of tasks typically consumes most resources. In traditional cloud workloads, studies such as the Google Cluster Traces and AWS workload analyses show that lightweight tasks often dominate, representing slightly more than 80$\%$ of total workloads. This allocation accounts for the growing dominance of IoT-driven and repetitive lightweight tasks while reserving the cloud for high-resource, complex workloads. The percentages can vary depending on specific applications and later in the document we perform analysis and Montecarlo simulations to model and measure the benefits for different split of workloads across end devices and the cloud.

\section{MATHEMATICAL FRAMEWORK}
\subsection{PARETO DISTRIBUTION}
We assume a Pareto distribution for workload sizes based on findings in \cite{barroso2013datacenter}, \cite{xu2017cloudworkload}, \cite{reiss2011googlecluster},\cite{dean2008mapreduce}, \cite{calheiros2011cloudsim}, \cite{baliga2011energy}, \cite{tucker2018energy}, and \cite{andrae2015electricity}.
\[
f(x; \alpha, x_m) = \frac{\alpha \, x_m^\alpha}{x^{\alpha + 1}}, \quad x \geq x_m
\]

\textbf{Where:}
\begin{itemize}
    \item $x$: Workload size.
    \item $\alpha = 2$: Shape parameter (controls skewness).
    \item $x_m = 1$: Minimum workload size.
\end{itemize}

\noindent The Pareto analysis demonstrates that the majority (~70$\%$-90$\%$) of workloads are lightweight and can run on existing compute resources without requiring specialized upgrades. For devices with constrained compute capabilities, nearby edge compute units such as smartphones, infotainment systems, or smart hubs can collaborate within a service mesh to offload and share workloads. This ensures that even microcontroller-based devices can contribute effectively to agentic workflows without prohibitive hardware investments \cite{intel2023edgenativeai}, \cite{deloitte2023aiworkloads}.\\\\
While some use cases may require upgrades from microcontrollers to more capable processors, advancements in AI model optimization such as quantization, pruning, and distillation are reducing the computational requirements for inference tasks. Edge-AI frameworks, such as TensorFlow Lite and ONNX Runtime, enable resource-limited devices to process AI workloads efficiently, maintaining a balance between performance and cost \cite{tensorflowlite}, \cite{turc2020mobilebert}, and \cite{arm2023cortex}.
\subsection{ENERGY AND COST MODELS}
\subsection*{Energy Consumption for Centralized Cloud}
\[
E_{\text{cloud}} = D_T \cdot (E_t + E_c)
\]
\textbf{Where:}
\begin{itemize}
    \item $E_{\text{cloud}}$: Total energy consumption in the centralized cloud.
    \item $D_T$: Total data volume generated annually (in GB/year).
    \item $E_t$: Energy per GB for data transmission to the cloud (in kWh/GB).
    \item $E_c$: Energy per GB for cloud processing (in kWh/GB).
\end{itemize}

\subsection*{Energy Consumption for Hybrid Edge Cloud}
\[
E_{\text{HEC}} = D_{\text{edge}} \cdot E_l + D_{\text{cloud}} \cdot (E_t + E_c)
\]
\textbf{Where:}
\begin{itemize}
    \item $E_{\text{HEC}}$: Total energy consumption for HEC.
    \item $D_{\text{edge}} = P_{\text{edge}} \cdot D_T$: Data processed locally on edge devices.
    \item $D_{\text{cloud}} = P_{\text{cloud}} \cdot D_T$: Data transferred to the cloud.
    \item $E_l$: Energy per GB for local processing on edge devices (0.5 kWh/GB).
    \item $P_{\text{edge}}$ and $P_{\text{cloud}}$: Probabilities of the workloads being assigned to the edge or cloud.
\end{itemize}

\subsection*{Cost Model for Centralized Cloud}
\[
C_{\text{cloud}} = D_T \cdot (C_b + C_h)
\]
\textbf{Where:}
\begin{itemize}
    \item $C_{\text{cloud}}$: Total cost in the centralized cloud.
    \item $C_b$: Bandwidth cost per GB.
    \item $C_h$: Hosting cost per GB.
\end{itemize}

\subsection*{Cost Model for Hybrid Edge Cloud}
\[
C_{\text{HEC}} = D_{\text{cloud}} \cdot (C_b + C_h) + D_{\text{edge}} \cdot C_s
\]
\textbf{Where:}
\begin{itemize}
    \item $C_{\text{HEC}}$: Total cost for HEC.
    \item $C_b$: Bandwidth cost per GB.
    \item $C_h$: Hosting cost per GB.
    \item $C_s$: Software cost per GB.
\end{itemize}

The savings is energy with HEC can then be described by the following simple formula:
\[
S_{\text{Energy}} = \frac{E_t + E_c - E_l}{E_t + E_c} \cdot P_{\text{edge}}
\]

Similarly, the cost savings can be expressed by:
\[
S_{\text{Cost}} = \frac{C_b + C_h - C_s}{C_b + C_h} \cdot P_{\text{edge}}
\]
These results assume a uniform workload split probability, meaning tasks are assigned randomly to the edge or cloud based on the split percentage.

\section{NUMERICAL ANALYSIS RESULTS}
\subsection{TRADITIONAL WORKLOADS (BASELINE)}

For an average of 2.4 GB per day as reported in many references, the total workload data processed is about 876 GB/year/device.\\

\noindent For our numerical analysis, we assume the following costs:

\subsubsection*{Energy Parameters}
\begin{itemize}
    \item $E_t$: Energy per GB for data transmission to the cloud: 0.7 kWh/GB.
    \item $E_c$: Energy per GB for cloud processing: 1.5 kWh/GB.
    \item $E_l$: Energy per GB for local processing: 0.5 kWh/GB.
\end{itemize}

\subsubsection*{Cost Parameters}
\begin{itemize}
    \item $C_b$: Bandwidth cost per GB for data transmission to the cloud (\$0.10/GB).
    \item $C_h$: Hosting costs per GB for cloud processing (\$0.20/GB).
    \item $C_s$: Software licensing costs for processing on end devices (\$0.02/GB, assume 10\% of cloud hosting costs).
\end{itemize}
Table 1 and Figure 1 show the costs and energy benefits of HEC compared to traditional cloud-based solutions.

\subsubsection*{Energy Costs:}
According to the numerical analysis, the centralized cloud energy consumption is about: 1,927 kWh/device/year. With HEC and assuming an 80$\%$ edge split, the energy consumption is approximately 674 kWh/device/year. The resulting savings with HEC is approximately: 65$\%$.
\subsection*{Bandwidth \& Hosting Cost:}
Without HEC, the centralized cloud bandwidth and hosting costs is on the average about $\$$263 per device per year for traditional workloads.\\\\
With HEC and assuming an 80$\%$ edge split, the total costs of hosting, bandwidth and software licenses is  $\$$66 per device per year. The resulting savings with HEC is about $\$$200 per device per year or approximately 75$\%$.
\subsection*{Saving Comparison with Agentic Workloads:}
For agentic workloads the percentage costs savings is the same and the energy saving is slightly less than traditional workloads (about 62$\%$) but the savings per device per year are almost an order of magnitude larger providing up to 10,000 kWh per year in energy and $\$$1,500 in bandwidth and hosting cost per year per device which emphasizes why HEC is indispensable to help make agentic economy sustainable.  \\

\noindent Agentic workflows, where devices autonomously host and run AI inference models, require scalable and energy-efficient processing capabilities. Although agentic workflows may sometimes require upgrades from microcontrollers to more capable processors, advancements in AI model optimization such as quantization, pruning, and distillation have enabled models like MobileBERT \cite{turc2020mobilebert}  and lightweight vision models to run efficiently on CPUs and NPUs. Additionally, edge collaboration via service mesh architectures ensures that smaller devices can offload tasks to nearby intelligent devices like smartphones or infotainment units in vehicles, reducing the need for specialized hardware investments \cite{tensorflowlite}, \cite{intel2023edgenativeai}.\\

\noindent The mimik platform’s device-first approach aligns perfectly with this need, enabling localized execution of AI inference models. Studies demonstrate that optimized AI models, such as MobileBERT and lightweight vision models, can run effectively on CPUs, NPUs, and resource-constrained devices without requiring GPUs \cite{turc2020mobilebert}.\\

\noindent For example, advancements in quantization, pruning, and distillation allow large AI models to be compressed and deployed on edge devices with minimal performance loss \cite{mit2016efficient}. Frameworks like TensorFlow Lite, ONNX Runtime, and Core ML further enable efficient inference on billions of existing devices ranging from PCs and smartphones to IoT sensors and smart cameras. These findings align with broader industry trends that show 70$\%$-90$\%$ of AI workloads are lightweight and can already be processed locally without GPUs. As AI frameworks continue to optimize for edge inference, even legacy devices will be capable of running narrow AI models effectively. While future hardware upgrades will enhance these capabilities, they are not a prerequisite for running most narrow language models and edge inference tasks \cite{tensorflowlite}, \cite{arm2023cortex}. \\\\\\\\

\begin{table}[ht]
    \centering
    \includegraphics[width=.95\linewidth]{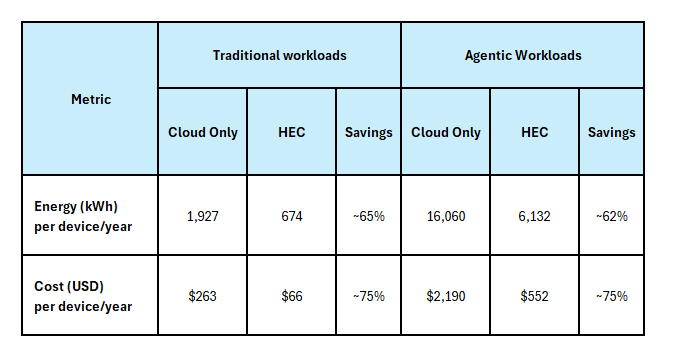}
    \caption{Energy and Cost Comparison of HEC with cloud-only solutions}
    \label{fig:enter-label}
\end{table}
\begin{figure}[ht]
    \centering
    \includegraphics[width=.95\linewidth]{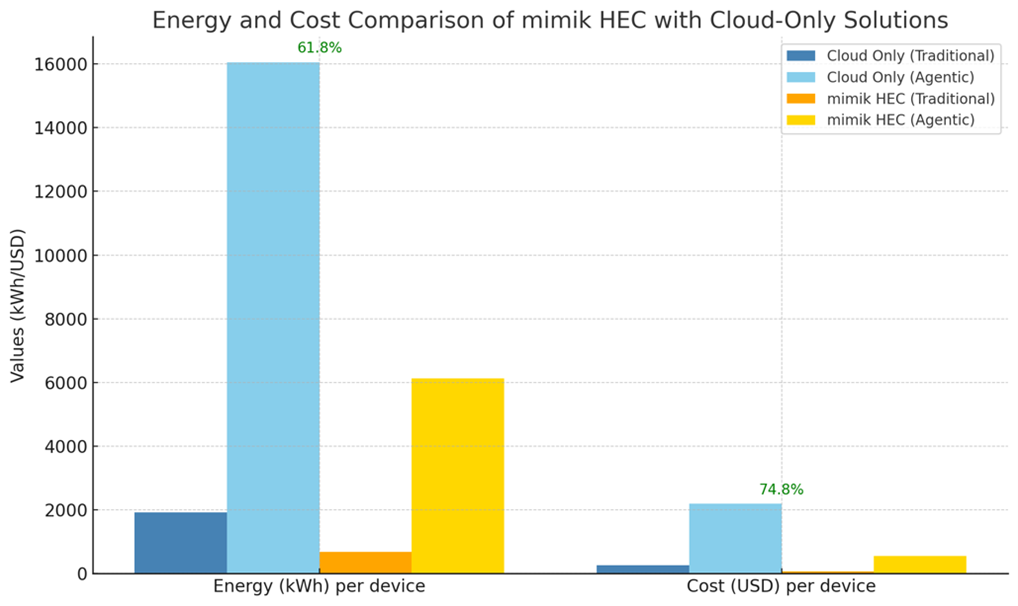}
    \caption{Energy and Cost Comparison of HEC with cloud-only solutions}
    \label{fig:enter-label-1}
\end{figure}

\begin{table}[ht]
    \centering
    \includegraphics[width=.95\linewidth]{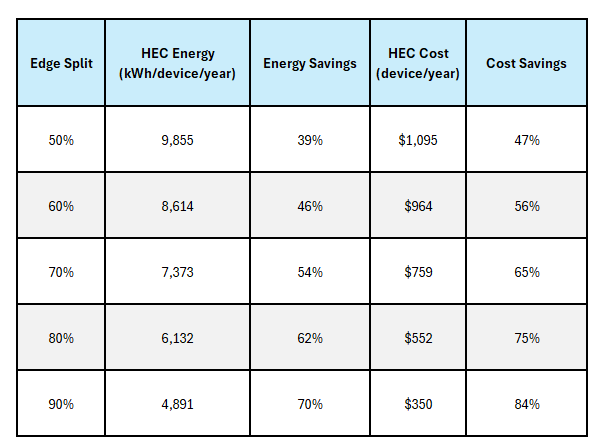}
    \caption{Energy and Cost Savings of HEC as a function of edge split.}
    \label{fig:enter-label-2}
\end{table}

\begin{figure}[ht]
    \centering
    \includegraphics[width=.95\linewidth]{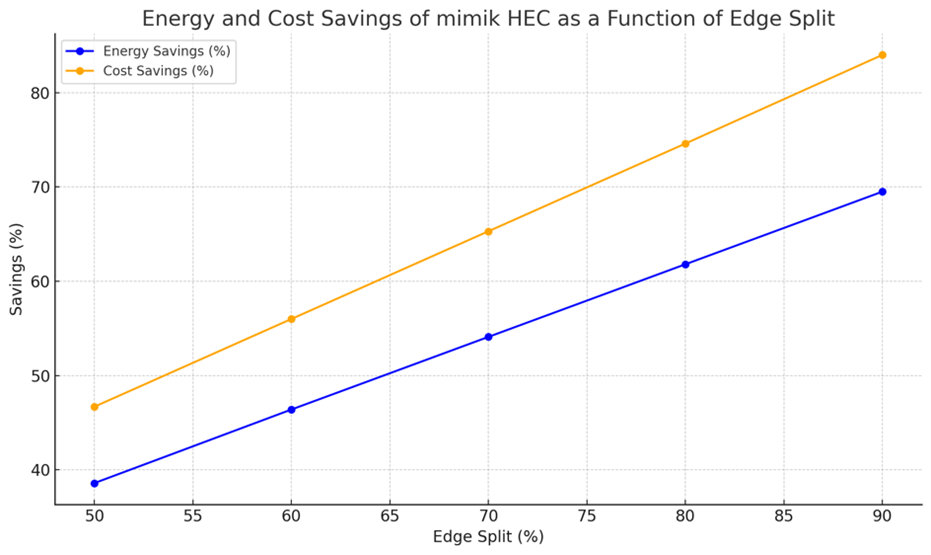}
    \caption{Energy and Cost Savings of HEC as a function of edge split.}
    \label{fig:enter-label-3}
\end{figure}
\noindent To consider the varying nature of workloads and their split across the cloud and edge, Table 2 and Figure 2 below show the results for the various edge split cases. As a reference the base only total energy per device for the cloud only case is 16,060 kWh per year and the cloud costs are $\$$2,190 per device per year.\\\\

\noindent The cost and energy benefits of HEC are realized even at lower edge splits. For instance, at an edge split of 30$\%$, energy and cost savings can still approach 25$\%$-30$\%$, providing meaningful reductions in operational expenses. This shows that HEC is economically viable even when edge splits are low. Incremental adoption allows organizations to realize immediate savings while gradually scaling edge processing over time. By reducing cloud transmission and compute costs, HEC remains superior to centralized models, particularly as workloads become more data-intensive \cite{deloitte2023aiworkloads}. \\\\
Importantly, in scenarios where edge processing is limited, HEC defaults to a centralized cloud model, ensuring no performance or cost penalties. This makes a device-first approach inherently superior, offering the flexibility to scale edge processing over time.


\section{SIMULATION RESULTS}
To validate the numerical results and model the real world more accurately, we can use Monte Carlo simulations. This  captures real-world variability in workloads distribution, making the results more realistic and actionable. Note that as in the numerical analysis, we incorporate an end device processing charge to include the cost of software licenses for mimik operating environment (mim-OE) in these simulations.\\
\\
The Monte Carlo simulations were performed under the following assumptions and parameters:

\begin{itemize}
    \item \textbf{Number of Devices:} Simulations were conducted for 10,000 devices and averaged.
    \item \textbf{Workload Generation:} Workload sizes were modeled using a Pareto distribution with a shape parameter $\alpha = 2$ and $\alpha = 3$, representing real-world workload variability. The minimum workload size was set at 1 GB per device. We assume a uniform workload split probability, meaning tasks are assigned randomly to the edge or cloud based on the split percentage.
    \item \textbf{Edge Split:} Simulations evaluated edge splits ranging from 50\% to 90\%, determining the proportion of workloads processed locally versus in the cloud.
\end{itemize}

\noindent The simulation results highlight energy and cost savings across edge splits ranging from 50\% to 90\%. The table 3 below presents the recalculated costs and savings, incorporating edge processing charges. The results for the two Pareto shape parameters were almost identical. While Pareto distributions differ in workload skewness, the normalization to a total annual workload of 7300 GB minimizes the effect of the skewness on overall energy and cost savings.

\begin{table}[ht]
    \centering
    \includegraphics[width=.95\linewidth]{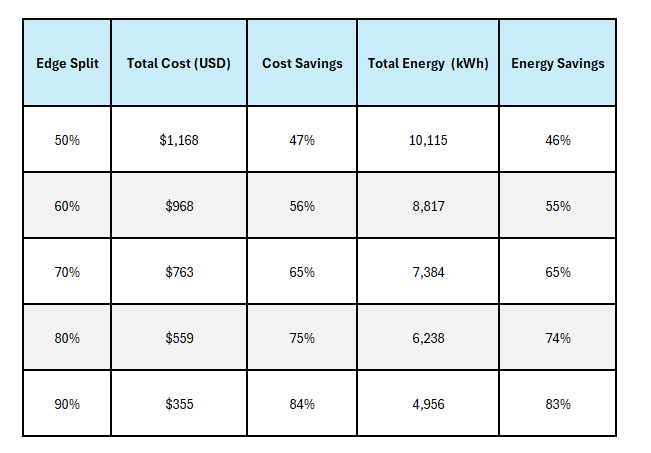}
    \caption{Energy and Cost Savings of HEC as a function of edge split with edge processing charges.}
    \label{fig:enter-label-4}
\end{table}

The simulation results provide several key insights into the benefits of HEC:

\begin{enumerate}
    \item \textbf{Cost Efficiency:} Even with a nominal charge for software licenses for device processing, HEC demonstrates significant cost savings compared to centralized cloud models. For example, at 80\% split, total costs are reduced by almost 75\%, highlighting the huge economic advantages of HEC.
    
    \item \textbf{Energy Efficiency:} HEC reduces the energy required for data transmission and centralized processing by enabling localized workloads. This not only reduces operational costs but also contributes to sustainability goals.
    
    \item \textbf{Scalability:} The simulations show that HEC scales effectively to handle large deployments, with savings increasing as more workloads are processed locally.
    
    \item \textbf{Real-World Implications:} The results emphasize the importance of device-first architectures for applications requiring low latency and high efficiency, such as IoT, AI, and autonomous systems.
\end{enumerate}

\section{Discussion}
\subsection{ENERGY EFFICIENCY}
The energy efficiency of HEC stems from its ability to process workloads locally on end devices, reducing the need for data transmission to and processing in centralized cloud data centers. This approach has a particularly profound impact in scenarios where data volumes are large and tasks are context-sensitive, such as in agentic workloads.\\
\\
\textbf{Traditional Workloads:} For traditional IoT workloads (e.g., smart devices generating 2.4 GB/day), the savings in energy consumption are as much as 80\%. A significant portion of traditional workloads involves lightweight, repetitive tasks that inherently consume less energy whether processed locally or centrally.
\\\\
\textbf{Agentic Workloads:} In contrast, agentic workloads, such as those generated by AI-driven applications (e.g., autonomous vehicles, drones, or robots), produce significantly more data—7,300 GB/year for 20 GB/day. These workloads consume far more energy when processed in the cloud due to both high transmission energy (5 kWh/GB) and computational energy (1.5 kWh/GB) for cloud processing. By processing \textbf{80\%} of these workloads locally on end devices, HEC reduces overall energy consumption by approximately \textbf{75\%}.
\\\\
The significantly higher energy savings for agentic workloads can be attributed to the exponential scaling of data transmission costs in centralized models. As workloads become more data-intensive, the relative efficiency of local processing becomes increasingly pronounced.
\subsection{COST REDUCTION}
The cost benefits of HEC are similarly striking, especially for data-intensive workloads. Cost savings are driven by reduced reliance on expensive bandwidth and cloud hosting services.\\\\
\textbf{Traditional Workloads:} For traditional workloads, the cost savings are significant even considering a nominal amount for the software. Most of the lightweight tasks are processed locally, drastically reducing bandwidth usage and hosting fees. While the overall data volume is smaller, the cost reduction still reflects the elimination of central cloud dependencies for most tasks.\\\\
\textbf{Agentic Workloads:} The per device cost savings for agentic workloads are almost an order of magnitude larger  than traditional workloads. This is because the proportion of data processed locally remains the same and the cost of transmitting and processing the remaining workloads in the cloud dominates centralized costs. The economic advantage becomes more substantial as data volume increases, making HEC an essential solution for high-scale AI-driven applications.

\section{Conclusion}
Our analysis and simulations demonstrate that HEC offers a transformative solution to the challenges of centralized cloud systems by providing a scalable, sustainable, and cost-effective architecture for managing modern workloads. By enabling localized processing for most tasks, HEC significantly reduces energy consumption and operational costs.\\

\noindent Compared to centralized cloud systems, which would consume over 16,000 kWh of energy and cost over \$2,000 annually for intelligent AI-enabled devices, HEC reduces energy consumption by approximately \textbf{10,000 kWh per device} and cuts costs by about \textbf{\$1,500 per year}.\\

\noindent These savings are particularly impactful for data-intensive applications such as autonomous vehicles, drones, and robotics, where real-time, context-aware processing is critical. For billions of devices operating worldwide, the total savings could amount to \textbf{tens of trillions of kWh annually and trillions of dollars} in cloud hosting and bandwidth costs. This scale of savings highlights the profound environmental and economic benefits of adopting HEC over centralized cloud architectures.\\\\
As we move into an era where tens of billions of interconnected devices will host AI agents, centralized cloud systems alone cannot meet this demand. The energy required to process and transmit data for billions of devices would be unsustainable, and the associated financial costs would far exceed current cloud resources' capabilities.  HEC addresses these resource limitations by leveraging existing device resources and reducing dependency on centralized infrastructure.\\\\
Advancements in AI model optimization such as quantization, pruning, and distillation enable existing devices to process most AI workloads efficiently without requiring specialized hardware upgrades. Frameworks like TensorFlow Lite and ONNX Runtime allow AI inference models to run effectively on CPUs and NPUs integrated into smartphones, PCs, and IoT devices. This means that the benefits of HEC and a device-first approach can be realized today, making it both practical and essential for supporting agentic workflows.\\\\
 HEC's scalability ensures that as billions of intelligent devices come online, existing compute resources are utilized efficiently. Optimized AI models and edge frameworks enable workloads to be processed locally, reducing reliance on cloud infrastructure while improving energy efficiency. This approach is critical for ensuring the feasibility and sustainability of AI-driven agentic workflows at scale.\\\\
Moreover, HEC aligns with global sustainability initiatives, such as net-zero emissions targets and the Paris Agreement, by enabling organizations to drastically reduce energy consumption and carbon footprints while scaling their operations. In a world where sustainability and scalability are paramount, HEC is not just advantageous. It is indispensable for supporting the intelligent systems of the future.

\nocite{*}
\bibliographystyle{unsrt}
\bibliography{bibfile}

\end{document}